\documentclass[prd,twocolumn,floatfix,preprintnumbers,altaffilletter]{revtex4}
\usepackage{graphicx,url,amssymb,amsmath,longtable,rotating,color,units,
wasysym,subfigure,epsfig}

\newcommand{\swift}{{\em Swift}}
\begin{document}

\author{%
J.~Aasi$^{1}$,
J.~Abadie$^{1}$,
B.~P.~Abbott$^{1}$,
R.~Abbott$^{1}$,
T.~Abbott$^{2}$,
M.~R.~Abernathy$^{1}$,
T.~Accadia$^{3}$,
F.~Acernese$^{4,5}$,
C.~Adams$^{6}$,
T.~Adams$^{7}$
% ...
}

\address {$^{1}$LIGO - California Institute of Technology, Pasadena, CA 91125, USA }
\address {$^{2}$Louisiana State University, Baton Rouge, LA 70803, USA }
\address {$^{3}$Laboratoire d'Annecy-le-Vieux de Physique des Particules (LAPP), Universit\'e de Savoie, CNRS/IN2P3, F-74941 Annecy-le-Vieux, France }
\address {$^{4}$INFN, Sezione di Napoli, Complesso Universitario di Monte S.Angelo, I-80126 Napoli, Italy }
\address {$^{5}$Universit\`a di Salerno, Fisciano, I-84084 Salerno, Italy }
\address {$^{6}$LIGO - Livingston Observatory, Livingston, LA 70754, USA }
\address {$^{7}$Cardiff University, Cardiff, CF24 3AA, United Kingdom }
% ...


\title{Search for long-lived gravitational-wave transients coincident with long gamma-ray bursts}

\date{\today}

\begin{abstract}
  Long gamma-ray bursts (GRBs) have been linked to extreme core-collapse supernovae from massive stars.
  Gravitational waves (GW) offer a probe of the physics behind long GRBs.
  We investigate models of long-lived ($\sim10$--$\unit[1000]{s}$) GW emission associated with the accretion disk of a collapsed star or with its protoneutron star remnant.
  Using data from LIGO's fifth science run, and GRB triggers from the \swift\ experiment, we perform a search for unmodeled long-lived GW transients.
  Finding no evidence of GW emission, we place $90\%$ confidence level upper limits on the GW fluence at Earth from long GRBs for three waveforms inspired by a model of GWs from accretion disk instabilities.
 These limits range from $F<\unit[3.5]{ergs \, cm^{-2}}$ to $F<\unit[1200]{ergs \, cm^{-2}}$, depending on the GRB and on the model, allowing us to probe optimistic scenarios of GW production out to distances as far as $\approx\unit[33]{Mpc}$.
  Advanced detectors are expected to achieve strain sensitivities $10\times$ better than initial LIGO, potentially allowing us to probe the engines of the nearest long GRBs.
\end{abstract}

\maketitle

\section{Introduction}\label{intro}
Gamma-ray bursts (GRBs) are divided into two classes~\citep{kouveliotou,gehrels}.
Short GRBs, lasting $\lesssim\unit[2]{s}$ and characterized by hard spectra, are thought to originate primarily from the merger of binary neutron stars or from the merger of a neutron star with a black hole~\cite{eichler,mochkovitch}.
On the other hand, long GRBs, lasting $\gtrsim\unit[2]{s}$ and characterized by soft spectra, are associated with the extreme core collapse of massive stars~\citep{campana,pian,soderberg,mazzali}.
In the standard scenario, long GRBs are the product of a relativistic outflow, driven either by a black hole with an accretion disk or a protomagnetar (see, e.g.,~\cite{thompson,macfadyen,woosley,pasczynski}).
At least two types of models have been proposed in which long GRBs may be associated with long-lived $\sim10$--$\unit[1000]{s}$ gravitational-wave (GW) transients.
One family of models relies on the formation of clumps in the accretion disk surrounding a newly formed black hole following core collapse~\cite{piro:07,vanputten:01,vanputten:08,bonnell,kiuchi}.
The motion of the clumps generates long-lived narrowband GWs.

The second family of models relies on GW emission from a nascent protoneutron star.
If the star is born spinning sufficiently rapidly~\cite{corsi}, or if it is spun up through fallback accretion~\cite{piro:11,pirothrane12}, it may undergo secular or dynamical instabilities~\cite{chandrasekhar,owen_rmodes}, which, in turn, are expected to produce long-lived narrowband GW transients~\cite{pirothrane12}.
Such rapidly spinning protoneutron stars have been invoked to help explain GRB afterglows~\cite{corsi}.

The goal of this work is to implement a search for generic long-lived GW transients coincident with long GRBs.
While we are motivated by the two families of models discussed above, we make only minimal assumptions about our signal: that it is long-lived and that it is narrowband, producing a narrow track on a frequency-time ($ft$)-map.

Our analysis builds on previous searches for GWs from GRBs by the LIGO~\cite{S5} and Virgo~\cite{virgo} detectors; (see more below).
However, this analysis differs significantly from previous LIGO-Virgo GRB analyses~\cite{grb:s5vsr1,grb-cbc:s5vsr1,grb051103,grb070201,grb:s6vsr23} since previous searches have focused on either short sub-second burst signals or modeled compact binary coalescence signals associated with short GRBs.
Here, however, we consider unmodeled signals lasting $\sim10$--$\unit[1000]{s}$ associated with the core-collapse death of massive stars.

During LIGO's fifth science run (S5) (Nov. 5, 2005--Sep. 30, 2007)~\cite{S5}, which provides the data for this analysis, GRBs were recorded by the \swift\ experiment~\cite{swift} at a rate of $\approx\unit[100]{yr^{-1}}$~\cite{gehrels:09}.
GRBs are most commonly detected at distances corresponding to redshifts $z\approx1$--$2$~\cite{gehrels:09}, though, nearby GRBs have been detected as close as $\unit[37]{Mpc}$~\citep{gal98}.
During S5, there were five nearby GRBs ($150$--$\unit[610]{Mpc}$)~\footnote{Here we assume a Hubble parameter $H_0=\unit[67.8]{km \, s^{-1} Mpc^{-1}}$ and matter energy density $\Omega_m=0.27$.}.
Unfortunately, LIGO was not observing at the time of these GRBs despite a coincident detector duty cycle of $\approx50\%$.
While none of the GRBs analyzed here are known to be nearby (having a luminosity distance $D_\text{luminosity}<\unit[1000]{Mpc}$ and redshift $\lesssim0.20$), the number of nearby GRBs during S5 bodes well for observing a nearby long GRB coincident with LIGO/Virgo data in the advanced detector era.

The remainder of this paper is organized as follows.
In Section~\ref{LIGO} we describe the LIGO observatories, in Section~\ref{method} we describe the methodology of our search, in Section~\ref{models} we describe the salient features of our signal model.
In Section~\ref{results} we describe our results and in Section~\ref{conclusions} we discuss implications and future work.

\section{The LIGO Observatories}\label{LIGO}
We analyze data from the $\unit[4]{km}$ H1 and L1 detectors in Hanford, WA and Livingston, LA respectively.
We use data from the S5 science run, during which LIGO achieved a strain sensitivity of  $\unit[\approx3\times10^{-23}]{Hz^{-1/2}}$ in the most sensitive band between $\sim100$--$\unit[200]{Hz}$~\cite{S5}.
The H1L1 detector pair provides the most sensitive data available during S5, though a multibaseline approach remains a future goal~\footnote{By adding additional detectors to the network (and computational complexity to the pipeline), it is possible to further improve sensitivity, though, for most GRBs, the gain for this analysis is expected to be marginal since the sensitivity is dominated by the most sensitive detector pair.}.

S5 saw a number of important milestones (see, e.g., \cite{crab-S5,S5stoch,geo}), but most relevant for our present discussion are results constraining the emission of GWs from GRBs~\cite{grb:s5vsr1,grb-cbc:s5vsr1,grb051103,grb070201} (see also~\cite{grb:s6vsr23}).
Previous results have limited the distance to long and short GRBs as a function of the available energy for generic waveforms~\cite{grb:s5vsr1,grb:s6vsr23} and also for compact binary coalescence waveforms~\cite{grb-cbc:s5vsr1}.
They have investigated the origin of two GRBs that might have occurred in nearby galaxies~\cite{grb051103,grb070201}.

Currently LIGO~\cite{aligo,indigo} and Virgo~\cite{virgo} observatories are undergoing major upgrades that are expected to lead to a factor of ten improvement in strain sensitivity, and thus distance reach.
The GEO detector~\cite{geo}, meanwhile, continues to take data while the KAGRA detector~\cite{kagra} is under construction.
This paper sets the stage for the analysis of long-lasting transients from GRBs in the advanced detector era and demonstrates a long-transient pipeline~\cite{stamp,stamp_glitch} that is expected to have more general applications~\cite{pirothrane12}.

\section{Method}\label{method}
We analyze GRB triggers---obtained through the Gamma-ray burst Coordinates Network~\cite{GCN} and consisting of trigger time, right ascension ($\text{RA}$), and declination ($\text{dec}$)---from the \swift\ satellite's Burst Alert Telescope, which has an angular resolution of $\approx0.02^\circ$--$0.07^\circ$~\cite{swiftpage} that is much smaller than the angular resolution of the GW detector network.
This resolution allows us to study GW frequencies up to $\unit[1200]{Hz}$ while neglecting complications from GRB sky localization errors; see~\cite{stamp_glitch}.

LIGO data are pre-processed to exclude corrupt and/or unusable data~\cite{lsc_glitch}.
In the frequency domain, we remove bins associated with highly non-stationary noise caused by known instrumental artifacts including $\unit[60]{Hz}$ harmonics and violin resonances~\cite{S5}.

We define a $\left[\unit[-600]{s},\unit[+900]{s}\right]$ {\em on-source} region around each GRB trigger.
The GW signal is assumed to exist only in the on-source region.
The $\unit[-600]{s}$ allows for possible delays between the formation of a compact remnant object and the emission of the gamma rays (see~\cite{grb:s6vsr23} and references therein).
The $\unit[+900]{s}$ is motivated by the hypothesis that GW production is related to GRB afterglows~\cite{corsi}, which can extend $\approx10$--$\unit[10^4]{s}$ after the initial GRB trigger, though most often the duration is $\lesssim\unit[1000]{s}$~\cite{Panaitescu}~\footnote{A search exploring times as long as $\unit[10^4]{s}$ after the GRB trigger presents additional computational burdens, and is therefore beyond our present scope.}.

Of the $131$ long ($t_{90}>\unit[2]{s}$) GRB triggers~\footnote{The $t_{90}$ time is defined as the duration in between the $5\%$ and $95\%$ total background-subtracted photon counts.} detected by the \swift\ satellite~\cite{swift} during S5, there are $29$ for which coincident H1L1 data are available for the entire $\unit[1500]{s}$ on-source region.
We analyze an additional $21$ GRB triggers for which $\geq\unit[1000]{s}$ of coincident H1L1 data are available (but not all $\unit[1500]{s}$) and hence searchable for signal, though, we do not include them in our upper-limit calculations described below.

We additionally require that the GRB is not located in a direction with poor network sensitivity, which can prevent the detection of even a loud signal (see the appendix for details).
Only one GRB is excluded on account of this requirement.

We consider a frequency range of $100$--$\unit[1200]{Hz}$, above which we cannot, at present, probe astrophysically interesting distances due to the increase in detector noise at high frequencies and the fact that strain amplitude falls like $1/f$ for a fixed energy budget.
Frequencies $\lesssim\unit[100]{Hz}$ are excluded since non-stationary noise in this band diminishes the sensitivity of the search; see~\cite{stamp_glitch}.

Following~\cite{stamp}, strain data from the $\unit[1100]{Hz}\times\unit[1500]{s}$ on-source region is converted to spectrograms ($ft$-maps) of strain cross- and auto-power spectra.
These $ft$-maps utilize Hann-windowed, $\unit[1]{s}$, $50\%$-overlapping segments with a frequency resolution of $\unit[1]{Hz}$ (see also~\cite{vanputten:01}).
The strain cross-power is given by~\cite{stamp}:
\begin{eqnarray}\label{eq:Y}
  \hat{Y}(t;f) & \equiv &
  \frac{2}{\cal N} \text{Re}\left[
    Q_{IJ}(t;f,\hat{\Omega}) \, 
    \tilde{s}_I^\star(t;f) \tilde{s}_J(t;f)
    \right] .
\end{eqnarray}
Here $t$ is the segment start time, $f$ is the frequency bin, ${\cal N}$ is a window normalization factor, $\hat\Omega$ is the search direction, and $\tilde{s}_I(t;f)$, $\tilde{s}_J(t;f)$ are discrete Fourier transforms of strain data for segment $t$ using detectors $I=\text{H1}$ and $J=\text{L1}$ respectively.
$Q_{IJ}(t;f,\hat\Omega)$ is a filter function, which takes into account the time delay between the detectors and their directional response; (see~\cite{stamp} for additional details).
The dependence of $\hat{Y}(t;f)$ on $\hat\Omega$ is implicit for the sake of notational compactness.
An estimator for the variance of $\hat{Y}(t;f)$ is given by~\cite{stamp}:
\begin{eqnarray}\label{eq:sigma}
  \hat\sigma^2(t;f) & \equiv & \frac{1}{2} \left|Q_{IJ}(t;f,\hat\Omega)
  \right|^2 P_I'(t;f) P_J'(t;f) ,
\end{eqnarray}
where $P_I'(t;f)$ and $P_J'(t;f)$ are the auto-powers measured in detectors $I$ and $J$, respectively and the prime denotes that they are calculated using the average of $n=8$ segments neighboring the one beginning at $t$ (four on each side).

\begin{figure*}[hbtp!]
  \begin{tabular}{cc}
    \psfig{file=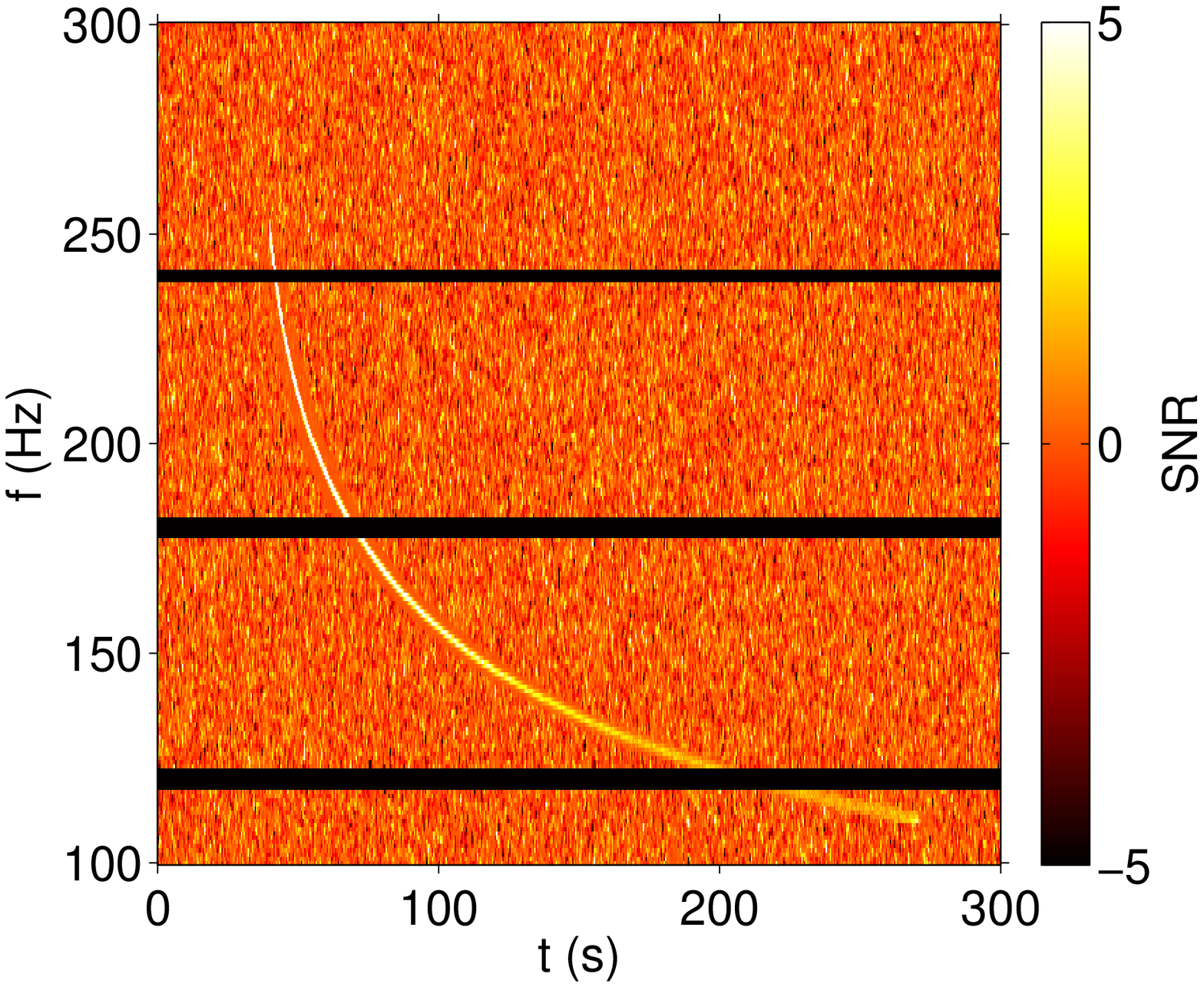,height=2.8in} & 
    \psfig{file=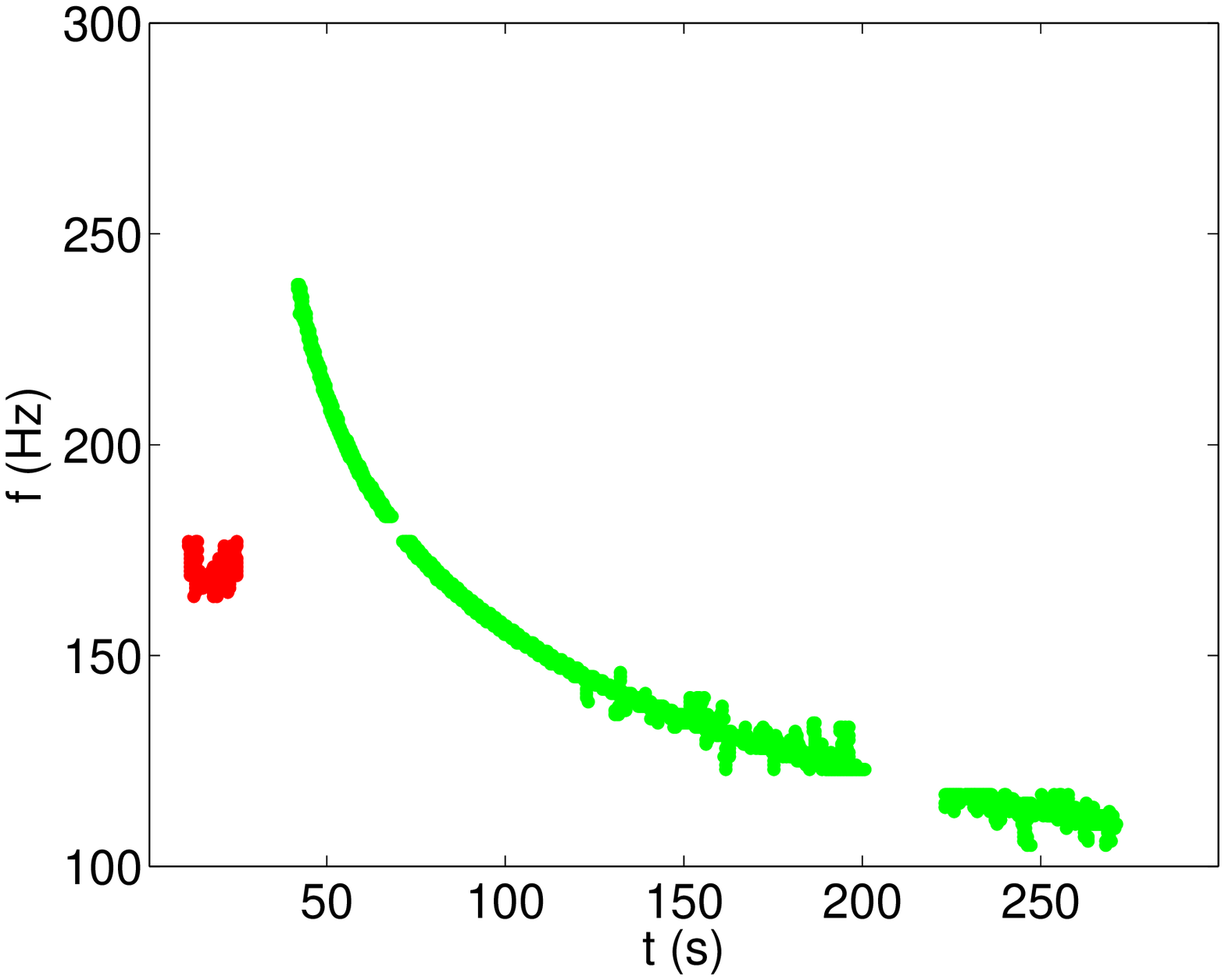,height=2.8in}
  \end{tabular}
  \caption{
    Recovery of a simulated waveform in time-shifted noise.  
    Left: $ft$-map of $\text{SNR}(t;f)$ for an injected accretion disk instability signal (model $c$).
    The horizontal black lines are removed frequency bins corresponding to instrumental artifacts.
    Right: significant clusters recovered with the clustering algorithm.
    Here the green cluster is due to injected GW signal while the red cluster is due to a noise fluctuation.
    In this example, the recovered $\text{SNR}_\text{tot}=290$ for the largest cluster (due to the GW signal) is well above the threshold of $30$ while noise fluctuations, such as the small red blob shown here, have typical recovered $\text{SNR}_\text{tot}=17$.
    (Note that the left-hand side color scale shows $\text{SNR}(t;f)$, defined in Eq.~\ref{eq:SNR} for each pixel, whereas $\text{SNR}_\text{tot}$, defined in Eq.~\ref{eq:snrtot}, is a property of a cluster consisting of many pixels.)
%
  }
  \label{fig:cluster}
\end{figure*}

Using Eqs.~\ref{eq:Y} and \ref{eq:sigma}, we cast our search for long GW transients as a pattern recognition problem (see Fig.~\ref{fig:cluster}).
GW signals create clusters of positive-valued pixels in $ft$-maps of signal-to-noise ratio:
\begin{equation}\label{eq:SNR}
  \text{SNR}(t;f) \equiv \hat{Y}(t;f)/\hat\sigma(t;f) ,
\end{equation}
whereas noise is randomly distributed with a mean of $\langle\text{SNR}(t;f)\rangle=0$.

We employ a track-search clustering algorithm for generic narrowband waveforms~\cite{burstegard}, which works by connecting $ft$-map pixels above a threshold and that fall within a fixed distance of nearby above-threshold pixels.
Clusters (denoted $\Gamma$) are ranked by the value of the total cluster signal-to-noise ratio $\text{SNR}_\text{tot}$:
\begin{equation}\label{eq:snrtot}
  \text{SNR}_\text{tot} = \frac{
  \sum_{t;f\in\Gamma} \hat{Y}(t;f) \, \hat\sigma^{-2}(t;f)
  }{
    \left( \sum_{t;f\in\Gamma} \hat\sigma^{-2}(t;f) \right)^{1/2}
  }
  .
\end{equation}
To evaluate the significance of the cluster with the highest $\text{SNR}_\text{tot}$ in the on-source region, we compare it to the background distribution, which is estimated using time-shifted data.

\begin{figure}[hbtp!]
  \psfig{file=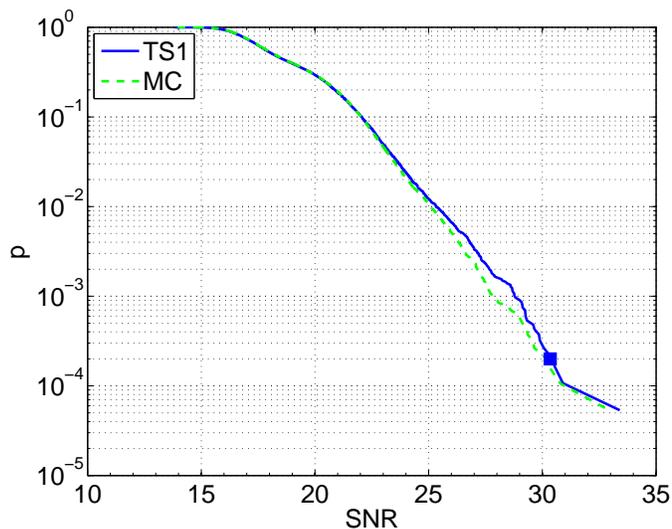,height=2.8in}
  \caption{
    Single-trigger false-alarm probability $p$ vs.\ $\text{SNR}_\text{tot}$ for time-shifted (TS) and Monte Carlo (MC) data.
    The marker indicates $\text{SNR}_\text{tot}^\text{th}$, the threshold for an interesting candidate for follow-up.
    $\text{SNR}_\text{tot}^\text{th}$ is defined such that the probability of observing any of the 50 GRB triggers with $\text{SNR}_\text{tot}>\text{SNR}_\text{tot}^\text{th}$ due to noise fluctuations is $<1\%$.
  }
  \label{fig:background}
\end{figure}

Time shifts, in which we offset the H1 and L1 strain series by an amount greater than the intersite GW travel time, provide a robust method of estimating background~\cite{s1_cbc}.
For each value of $\text{SNR}_\text{tot}$ we assign a false-alarm probability $p$ by performing many trials with time-shifted data (see Fig.~\ref{fig:background}).
The false-alarm probability for $\text{SNR}'_\text{tot}$ is given by the fraction of time-shifted trials for which we observed $\text{SNR}_\text{tot}\geq\text{SNR}'_\text{tot}$.
We apply a noise transient identification algorithm~\cite{stamp_glitch} in order to mitigate contamination from non-stationary noise.
Similar consistency-check noise transient identification is performed in previous searches for unmodeled GW, e.g.,~\cite{grb:s5vsr1}.
The relatively good agreement in Fig.~\ref{fig:background} between time-shifted and Monte Carlo data (colored Gaussian strain noise) is attributable in part to the stability of LIGO strain noise for frequencies $>\unit[100]{Hz}$ on long time scales~\cite{stamp_glitch}.

Using time-shifted data, we determine the interesting-candidate threshold $\text{SNR}_\text{tot}^\text{th}$ such that the probability of observing any of the 50 GRB triggers with $\text{SNR}_\text{tot}>\text{SNR}_\text{tot}^\text{th}$ due to noise fluctuations is $< 1\%$.
We find that the threshold for an interesting candidate is $\text{SNR}_\text{tot}^\text{th}=30$.
Interesting candidates, if they are observed, are subjected to further study.

\begin{figure}[hbtp!]
  \psfig{file=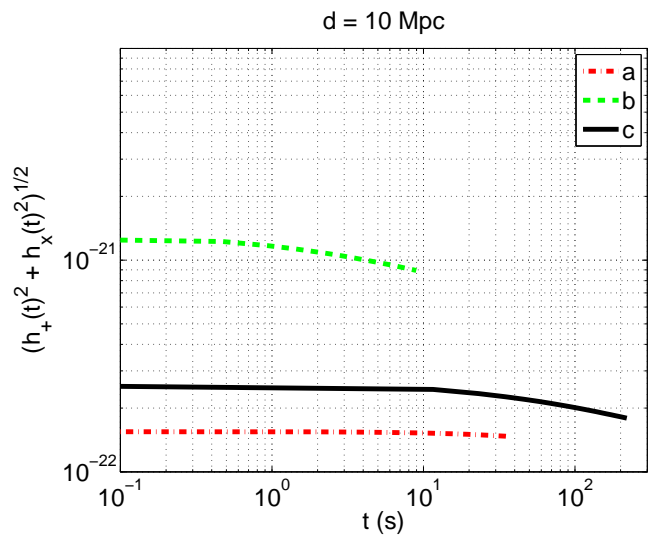,height=2.8in}
  \caption{
    Strain amplitude $[h_+^2(t)+h_\times^2(t)]^{1/2}$ vs.\ time for the waveforms in Table~\ref{tab:model1} assuming a reference distance of $\unit[10]{Mpc}$.
    }
  \label{fig:hphx}
\end{figure}

\section{Signal models}\label{models}
In order to constrain physical parameters such as fluence in the absence of a GW detection, it is necessary to have a waveform model.
In cases where there is no trusted waveform, one must employ a toy model which is believed to encompass the salient features of the astrophysical phenomenology, such as the sine-Gaussians used in short GW burst analyses~\cite{grb:s6vsr23}.

For our toy model, we employ accretion disk instability (ADI) waveforms~\cite{lucia} (based on~\cite{vanputten:01,vanputten:08} and references therein) in which a spinning black hole of mass $M$ (with typical values $3 M_\odot-10 M_\odot$) drives turbulence in an accretion torus of mass $m\approx1.5 M_\odot$.
This turbulence causes the formation of clumps of mass $\epsilon m$ (with typical values $0.015 M_\odot-0.3 M_\odot$), the motion of which emits GWs.
In optimistic models, as much as $E_\text{GW}=0.1 M_\odot c^2$ is emitted in GWs~\cite{vanputten:01}.
We emphasize that, like the sine-Gaussian waveforms used in short GW burst analyses, these waveforms should be taken as toy model representations of a GW signal for which there is significant theoretical uncertainty.

The model is additionally parameterized by a dimensionless spin parameter $a^\star\equiv(c/G)J_\text{BH}/M^2$, bounded by $[0,1)$, where $J_\text{BH}$ is the angular momentum of the black hole~\cite{lucia}.
An $ft$-map of $\text{SNR}(t;f)$ illustrating an injected ADI waveform with parameters $M=10M_\odot$, $m=1.5M_\odot$, $\epsilon=0.04$ and $a^\star=0.95$ (model $c$) is shown in the left-hand panel of Fig.~\ref{fig:cluster}.
(The GW frequency decreases with time as the black hole spins down and the innermost stable circular orbit changes.)
The waveforms are calculated assuming a circularly polarized source (inclination angle $\iota\approx0$), which is a reasonable assumption given that long GRBs are thought to be observed almost parallel to the angular momentum vector~\cite{gal-yam,racusin}.

We utilize different combinations of parameters to create three waveforms (denoted $a$, $b$, and $c$), which are summarized in Table~\ref{tab:model1} and Fig.~\ref{fig:hphx}.
By varying the model parameters, we obtain signals of varying durations ($9$--$\unit[231]{s}$).
For these three waveforms we constrain GW fluence---the GW energy flowing through a unit area at the detector integrated over the emission time.
The fluence is defined as:
\begin{equation}\label{eq:fluence}
  F \equiv \frac{c^3}{16\pi G} \int dt \left(
  \dot{h}_+^2(t) + \dot{h}_\times^2(t) \right) .
\end{equation}

By assuming a fixed GW energy budget $E_\text{GW}=0.1 M_\odot c^2$, it is possible to cast the fluence limits as limits on the distance $D$ to the GRB.
The relationship between fluence, distance, and energy is given by
\begin{equation}\label{eq:distance}
  D = \left( \frac{5}{2}\frac{E_\text{GW}}{4\pi F} \right)^{1/2} .
\end{equation}
The factor of $5/2$ arises from the assumption that the source emits face-on, which causes modest enhancement in observed fluence compared to a source observed edge-on.

\begin{table}
\begin{tabular}{|c|c|c|c|c|c|c|c|}
  \hline
  {\bf ID} & {\bf $M$} & {\bf $a^\star$} & {\bf $\epsilon$} & {\bf $m$}
  & {\bf $t_\text{dur}$} & {\bf $f$ (Hz)}\\
  \hline
  a & 5 & $0.3$ & $0.05$ & $1.5$ & $39$ & 131--171\\
  \hline
  b & 10 & $0.95$ & $0.2$ & $1.5$ & $9$ & 90--284\\
  \hline
  c & 10 & $0.95$ & $0.04$ & $1.5$ & $231$ & 105--259\\
  \hline
\end{tabular}
  \caption{
    Parameters for waveforms~\cite{lucia} inspired by~\cite{vanputten:01,vanputten:08}.
    $M$ is black hole mass in units of $M_\odot$, $a^\star$ is a dimensionless spin parameter, $\epsilon$ is the fraction of torus mass that clumps, and $m$ is the torus mass in units of  $M_\odot$.
    The free parameters ($M$, $a^\star$, $\epsilon$, $m$) are selected within the range of expected values in order to produce a range of signal durations.
  }
  \label{tab:model1}
\end{table}

\section{results}\label{results}
Properties of the loudest cluster for each GRB trigger including its signal-to-noise ratio $\text{SNR}_\text{tot}$ and its false-alarm probability $p$ are given in Table~\ref{tab:grb}.
Of the $50$ GRB triggers analyzed in this study, the most significant was GRB~070621 with $\text{SNR}_\text{tot}=24$ corresponding to a single-trigger false-alarm probability of $p=2.3\%$.
The probability of observing $\text{SNR}_\text{tot}\geq24$ among our $50$ GRB triggers is $69\%$.

Since we find no evidence of long-lived GW transients, we set $90\%$ confidence level (CL) upper limits on the GW fluence for each GRB trigger for the three test models considered.
To calculate these limits we perform pseudo experiments in which we inject waveforms $a$, $b$, and $c$.
All three waveforms are normalized to a fixed energy budget $E_\text{GW}=0.1 M_\odot$ by multiplying each strain time series by a constant~\footnote{In reality, $E_\text{GW}$ depends on model parameters, but it is useful for our present purposes to normalize all three waveforms to the same energy in order to observe how sensitivity varies with signal duration and morphology.}.
We vary the distance to the source in order to determine the distance for which $90\%$ of the injected signals are recovered with an $\text{SNR}_\text{tot}$ exceeding the loudest cluster in the on-source region.
From these distance limits, we obtain fluence limits from Eq.~\ref{eq:distance}.

GW strain measurements are subject to systematic calibration uncertainties.
For S5 H1,L1 and for $f<\unit[2000]{Hz}$, this error is estimated to be $10.4\%,14.4\%$ in amplitude~\cite{S5calib}.
In order to take calibration error into account in our upper limit calculation, we assume the true fluence is some number $\lambda$ times the measured fluence, and that $\lambda$ is Gaussian distributed with a mean of $1$ and a width of $\sqrt{10.4\%^2 + 14.4\%^2}=17.8\%$.
Marginalizing over $\lambda$ leads to a $15\%$ reduction in our distance sensitivity.
Phase and timing calibration errors are negligible for this analysis~\cite{S5calib}.

The $90\%$ CL limits for models $a,b,$ and $c$ are reported in Table~\ref{tab:ul}.
We report upper limits on fluence and lower limits on distance assuming a GW energy budget of $E_\text{GW}=0.1 M_\odot c^2$.
For model $a$, we place upper limits on GW fluence of $3.5$--$\unit[1000]{ergs \, cm^{-2}}$ (corresponding to distance lower limits of $1.9$--$\unit[33]{Mpc}$).
For model $b$, the corresponding limits are $F<4.4$--$\unit[410]{ergs \, cm^{-2}}$ ($D>3.0$--$\unit[29]{Mpc}$), and for model $c$, $F<16$--$\unit[1200]{ergs \, cm^{-2}}$ ($D>1.8$--$\unit[15]{Mpc}$).
The variation in limits for a given model is due primarily to the direction-dependent antenna response factors, which cause $\sigma(t;f)$ to vary by two orders of magnitude for different search directions.
The GRB for which we set the best limits is GRB~070611 while the least sensitive limits are placed on GRB~070107.

Given a fixed waveform with an overall normalization constant, fluence limits are proportional to limits on (the square) of the root-sum-squared strain
\begin{equation}
  h_\text{rss}^2\equiv\int dt \left( h_+^2(t) + h_\times^2(t)\right)
  = k F ,
\end{equation}
where $k$ is a waveform-dependent constant.
Using this relation, we can alternatively present the limits as 
\begin{equation}
  \begin{split}
    h^a_\text{rss}<7.0\times10^{-22}(F/\unit[3.5]{erg \, cm^{-2}})^{1/2} \\
    h^b_\text{rss}<7.7\times10^{-22}(F/\unit[4.4]{erg \, cm^{-2}})^{1/2} \\
    h^c_\text{rss}<1.5\times10^{-21}(F/\unit[16]{erg \, cm^{-2}})^{1/2}
    \end{split}
\end{equation}
The superscript of $h_\text{rss}$ refers to the different models.


\section{Implications and future work}\label{conclusions}
In the most optimistic scenarios for the production of GWs in stellar collapse, it has been claimed that as much as $E_\text{GW}=0.1 M_\odot c^2$ of energy is converted into GWs~\cite{vanputten:01}.
The GW signature from the actual core collapse, as opposed to subsequent emission from an accretion disk or from a protoneutron star remnant, is expected to be significantly less energetic, with a typical energy budget of $E_\text{GW}=\sim10^{-11}$--$10^{-7}M_\odot$~\cite{ott:11}.

\begin{figure}[hbtp!]
  \psfig{file=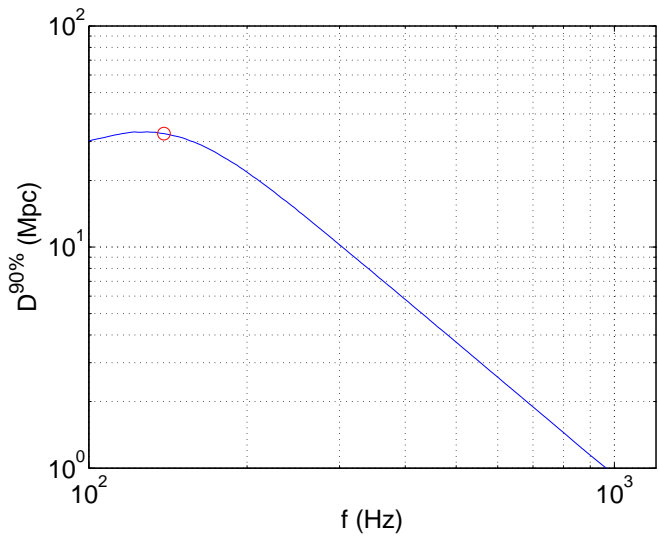,height=2.8in}
  \caption{
    Best-case-scenario, approximate $90\%$ CL lower limits on the distance to sources of long-lived GW transients extrapolated from fluence upper limits.
    Following~\cite{vanputten:01}, we assume a GW energy budget of $0.1 M_\odot c^2$.
    The limits are calculated using the best fluence constraints from Table~\ref{tab:ul}: $F^{90\%}=\unit[3.5]{ergs \, cm^{-2}}$ from GRB070611, model $a$.
    The GW power spectral peak is marked with a red circle.
    The frequency dependence should be taken as indicating an approximate trend as our waveforms are not monochromatic.
    Also, the extrapolation assumes a smooth detector noise curve, which only approximates the LIGO detector noise.
  }
  \label{fig:distances}
\end{figure}

By comparing our best fluence upper limits $F^{90\%}=\unit[3.5]{ergs \, cm^{-2}}$ (GRB070611, model $a$) with this prediction, we extrapolate approximate distance lower limits as a function of frequency for this best-case scenario; see Fig.~\ref{fig:distances}.
In the most sensitive frequency range between $100$--$\unit[200]{Hz}$, the limits are as large as $D^{90\%}=\unit[33]{Mpc}$.
They fall like $D^{90\%}\propto f^{-2}$ above $\unit[200]{Hz}$ due to increasing the detector shot noise as well as from the relationship between energy and strain $E_\text{GW}\propto f^2 h^2$.
The limits in Fig.~\ref{fig:distances} scale like $D^{90\%}\propto E_\text{GW}^{1/2}$ and $D^{90\%}\propto (F^{90\%})^{-1/2}$.
The GW power spectral peak frequency is marked with a red circle.
Note that the waveforms we consider here are not characterized by a single frequency, and so Fig.~\ref{fig:distances} should be taken as an approximate indicator of how results scale with frequency.

If the GW frequency is high ($f\gtrsim\unit[1]{kHz}$)~\cite{vanputten:01}, the reach of initial LIGO is only $\lesssim\unit[1]{Mpc}$ due to the fact that distance sensitivity falls off rapidly with frequency: $D\propto f^{-2}$.
The nearest GRB in our set with a known redshift measurement, GRB~070420, is estimated to have occurred at $z=0.48$--$0.93$ ($D_\text{luminosity}=2800$--$\unit[6400]{Mpc}$)~\cite{swift_redshifts}, well beyond our exclusion distances even for lower-frequency emission.

While we are therefore unable to rule out the most extreme models of GW emission with the present analysis, we have demonstrated that initial LIGO can test optimistic models out to distances as far as $\approx\unit[33]{Mpc}$ depending on the GW frequency and the detector orientation during the time of the GRB.
Advanced LIGO and Advanced Virgo are expected to achieve strain sensitivities $10\times$ better than the initial LIGO data analyzed here, which will be sufficient to test extreme models out to $D\approx\unit[330]{Mpc}$.
As discussed in Section~\ref{intro}, GRBs are not infrequent at such distances~\footnote{Given a GRB rate of $\approx\unit[0.5]{Gpc^{-3}yr^{-1}}$ and a detection volume of $(4\pi/3) (\unit[0.33]{Gpc})^3$, we expect an event rate of $\approx\unit[0.1]{yr^{-1}}$, though the local rate is actually somewhat higher.}.

Meanwhile, work is ongoing to develop more sophisticated data analysis procedures, to further enhance sensitivity.
By tuning our analysis pipeline~\cite{stamp} for long-lived signals, we estimate that we can detect ADI waveforms for sources that are twice as distant as could have been detected by previous searches tuned for short signals~\cite{grb:s5vsr1,grb:s6vsr23} (corresponding to an increase in detection volume of $\approx8\times$).
In order to achieve additional improvements in sensitivity, work is ongoing to explore alternative pattern recognition strategies that relax the requirement that $\text{SNR}(t;f)$ exceeds some threshold to form a pixel cluster (see Eq.~\ref{eq:SNR}).

Long GRBs are by no means the only interesting source of long GW transients.
In~\cite{piro:11,pirothrane12} it was argued that core-collapse supernovae can trigger the production of long-lived GW emission through fallback accretion.
While the predicted strains are much less than the most extreme models considered here, the local rate of supernovae is much higher than the local rate of long GRBs, and preliminary sensitivity estimates suggest that fallback accretion-powered signals are interesting targets for Advanced LIGO/Virgo~\cite{pirothrane12}.
Other scenarios for long-lived GW production explored in~\cite{stamp}, including protoneutron star convection and eccentric black hole binaries, remain areas of investigation.
This analysis paves the way for future studies probing unmodeled long-lived GW emission.

\begin{acknowledgements}
The authors gratefully acknowledge the support of the United States
National Science Foundation for the construction and operation of the
LIGO Laboratory, the Science and Technology Facilities Council of the
United Kingdom, the Max-Planck-Society, and the State of
Niedersachsen/Germany for support of the construction and operation of
the GEO600 detector, and the Italian Istituto Nazionale di Fisica
Nucleare and the French Centre National de la Recherche Scientifique
for the construction and operation of the Virgo detector. The authors
also gratefully acknowledge the support of the research by these
agencies and by the Australian Research Council, 
the International Science Linkages program of the Commonwealth of Australia,
the Council of Scientific and Industrial Research of India, 
the Istituto Nazionale di Fisica Nucleare of Italy, 
the Spanish Ministerio de Econom\'ia y Competitividad,
the Conselleria d'Economia Hisenda i Innovaci\'o of the
Govern de les Illes Balears, the Foundation for Fundamental Research
on Matter supported by the Netherlands Organisation for Scientific Research, 
the Polish Ministry of Science and Higher Education, the FOCUS
Programme of Foundation for Polish Science,
the Royal Society, the Scottish Funding Council, the
Scottish Universities Physics Alliance, The National Aeronautics and
Space Administration, 
OTKA of Hungary,
the Lyon Institute of Origins (LIO),
the National Research Foundation of Korea,
Industry Canada and the Province of Ontario through the Ministry of Economic Development and Innovation, 
the National Science and Engineering Research Council Canada,
the Carnegie Trust, the Leverhulme Trust, the
David and Lucile Packard Foundation, the Research Corporation, and
the Alfred P. Sloan Foundation.
\end{acknowledgements}

\begin{appendix}

\begin{table*}
\scalebox{0.9}{ 
  \begin{tabular}{|l|c|c|c|c|c|c|c|c|c|}
    \hline
    & {\bf GRB} & {\bf GPS} & {\bf RA} (hr) & {\bf DEC} (deg) 
    & ${\bf t_{90} (s)}$ & {\bf All data?} & 
    ${\bf \text{SNR}_\text{tot}}$ & {\bf $p$ (\%)} \\\hline
   1 & GRB060116 & 821435861 & 5.65 & -5.45 & 105.9 & no& 16 & 89.1 \\\hline
    2 & GRB060322 & 827103635 & 18.28 & -36.82 & 221.5 & no& 18 & 49.7 \\\hline
    3 & GRB060424 & 829887393 & 0.49 & 36.79 & 37.5 & yes& 20 & 27.5 \\\hline
    4 & GRB060427 & 830173404 & 8.28 & 62.65 & 64.0 & no& 16 & 87.5 \\\hline
    5 & GRB060428B & 830249692 & 15.69 & 62.03 & 57.9 & yes& 16 & 85.8 \\\hline
    6 & GRB060510B & 831284548 & 15.95 & 78.60 & 275.2 & no& 18 & 55.5 \\\hline
    7 & GRB060515 & 831695286 & 8.49 & 73.56 & 52.0 & no& 18 & 59.1 \\\hline
    8 & GRB060516 & 831797028 & 4.74 & -18.10 & 161.6 & no& 17 & 84.9 \\\hline
    9 & GRB060607B & 833758378 & 2.80 & 14.75 & 31.1 & yes& 21 & 21.4 \\\hline
    10 & GRB060707 & 836343033 & 23.80 & -17.91 & 66.2 & no& 18 & 46.9 \\\hline
    11 & GRB060714 & 836925134 & 15.19 & -6.54 & 115.0 & no& 24 & 3.1 \\\hline
    12 & GRB060719 & 837327050 & 1.23 & -48.38 & 66.9 & yes& 17 & 70.0 \\\hline
    13 & GRB060804 & 838711713 & 7.48 & -27.23 & 17.8 & no& 18 & 55.2 \\\hline
    14 & GRB060807 & 838996909 & 16.83 & 31.60 & 54.0 & no& 22 & 10.1 \\\hline
    15 & GRB060813 & 839544636 & 7.46 & -29.84 & 16.1 & yes& 18 & 50.7 \\\hline
    16 & GRB060814 & 839631753 & 14.76 & 20.59 & 145.3 & no& 21 & 16.7 \\\hline
    17 & GRB060908 & 841741056 & 2.12 & 0.33 & 19.3 & yes& 19 & 43.4 \\\hline
    18 & GRB060919 & 842687332 & 18.46 & -50.99 & 9.1 & no& 19 & 43.2 \\\hline
    19 & GRB060923B & 843046700 & 15.88 & -30.91 & 8.6 & no& 21 & 18.3 \\\hline
    20 & GRB061007 & 844250902 & 3.09 & -50.50 & 75.3 & yes& 18 & 47.2 \\\hline
    21 & GRB061021 & 845480361 & 9.68 & -21.95 & 46.2 & yes& 20 & 30.9 \\\hline
    22 & GRB061102 & 846464445 & 9.89 & -17.00 & 45.6 & no& 19 & 43.7 \\\hline
    23 & GRB061126 & 848566090 & 5.77 & 64.20 & 70.8 & no& 22 & 7.9 \\\hline
    24 & GRB061202 & 849082318 & 7.01 & -74.59 & 91.2 & yes& 20 & 32.2 \\\hline
    25 & GRB061218 & 850449919 & 9.95 & -35.22 & 6.5 & no& 17 & 75.8 \\\hline
    26 & GRB061222B & 850795876 & 7.02 & -25.86 & 40.0 & yes& 21 & 16.6 \\\hline
    27 & GRB070107 & 852206732 & 10.63 & -53.20 & 347.3 & yes& 18 & 53.4 \\\hline
    28 & GRB070110 & 852448975 & 0.06 & -52.98 & 88.4 & yes& 16 & 88.4 \\\hline
    29 & GRB070208 & 854961048 & 13.19 & 61.95 & 47.7 & yes& 22 & 13.0 \\\hline
    30 & GRB070219 & 855882630 & 17.35 & 69.34 & 16.6 & yes& 17 & 75.6 \\\hline
    31 & GRB070223 & 856228514 & 10.23 & 43.13 & 88.5 & yes& 18 & 58.2 \\\hline
    32 & GRB070318 & 858238150 & 3.23 & -42.95 & 74.6 & yes& 20 & 24.9 \\\hline
    33 & GRB070330 & 859330305 & 17.97 & -63.80 & 9.0 & yes& 16 & 95.7 \\\hline
    34 & GRB070412 & 860376437 & 12.10 & 40.13 & 33.8 & yes& 17 & 74.8 \\\hline
    35 & GRB070420 & 861085107 & 8.08 & -45.56 & 76.5 & yes& 20 & 27.5 \\\hline
    36 & GRB070427 & 861697882 & 1.92 & -27.60 & 11.1 & yes& 19 & 40.0 \\\hline
    37 & GRB070506 & 862464972 & 23.15 & 10.71 & 4.3 & yes& 21 & 20.6 \\\hline
    38 & GRB070508 & 862633111 & 20.86 & -78.38 & 20.9 & yes& 18 & 48.9 \\\hline
    39 & GRB070509 & 862714121 & 15.86 & -78.66 & 7.7 & yes& 17 & 84.6 \\\hline
    40 & GRB070520B & 863718307 & 8.13 & 57.59 & 65.8 & no& 18 & 55.7 \\\hline
    41 & GRB070529 & 864478122 & 18.92 & 20.65 & 109.2 & yes& 17 & 78.6 \\\hline
    42 & GRB070611 & 865562247 & 0.13 & -29.76 & 12.2 & yes& 22 & 7.9 \\\hline
    43 & GRB070612B & 865664491 & 17.45 & -8.75 & 13.5 & no& 17 & 77.5 \\\hline
    44 & GRB070621 & 866503073 & 21.59 & -24.81 & 33.3 & yes& 24 & 2.3 \\\hline
    45 & GRB070714B & 868424383 & 3.86 & 28.29 & 64.0 & yes& 19 & 38.8 \\\hline
    46 & GRB070721B & 869049242 & 2.21 & -2.20 & 340.0 & yes& 20 & 33.3 \\\hline
    47 & GRB070805 & 870378959 & 16.34 & -59.96 & 31.0 & yes& 17 & 82.0 \\\hline
    48 & GRB070911 & 873525478 & 1.72 & -33.48 & 162.0 & no& 18 & 59.4 \\\hline
    49 & GRB070917 & 874049650 & 19.59 & 2.42 & 7.3 & no& 19 & 37.7 \\\hline
    50 & GRB070920B & 874357486 & 0.01 & -34.84 & 20.2 & no& 22 & 8.4 \\\hline
  \end{tabular}
}
  \caption{
    Swift long GRB triggers coincident with S5 H1L1 data and associated GW search results.
    ``All data?'' asks whether there is coincident LIGO data for all $\unit[1500]{s}$ in the on-source region (yes) or for just some of it (no).
    $\text{SNR}_\text{tot}$ is the signal-to-noise ratio for the loudest cluster and $p$ is the single-trial false alarm probability.
  }
  \label{tab:grb}
\end{table*}

\begin{table*}
  \begin{tabular}{|c|c|c|c||c|c|c|}
    \hline
    \multicolumn{1}{|c|}{} &
    \multicolumn{3}{c||}{90\% UL on $F$ ($\unit[]{ergs \, cm^{-2}}$)} &
    \multicolumn{3}{c|}{90\% LL on $D$ ($\unit[]{Mpc}$)} \\
    \multicolumn{1}{|c|}{ID} & 
    \multicolumn{3}{c||}{model} &
    \multicolumn{3}{c|}{model} \\
    & a & b & c & a & b & c \\\hline
GRB060424 & 20 & 25 & 71 & 14 & 12 & 7.2 \\\hline
GRB060428B & 4.9 & 8.7 & 21 & 28 & 21 & 13 \\\hline
GRB060607B & 94 & 120 & 330 & 6.3 & 5.7 & 3.3 \\\hline
GRB060719 & 20 & 30 & 71 & 14 & 11 & 7.3 \\\hline
GRB060813 & 25 & 26 & 74 & 12 & 12 & 7.1 \\\hline
GRB060908 & 5.6 & 6.9 & 20 & 26 & 23 & 14 \\\hline
GRB061007 & 120 & 180 & 620 & 5.6 & 4.6 & 2.5 \\\hline
GRB061021 & 21 & 26 & 89 & 13 & 12 & 6.5 \\\hline
GRB061202 & 15 & 22 & 64 & 16 & 13 & 7.7 \\\hline
GRB061222B & 1000 & 80 & 280 & 1.9 & 6.8 & 3.7 \\\hline
GRB070107 & 270 & 410 & 1200 & 3.7 & 3.0 & 1.8 \\\hline
GRB070110 & 6.7 & 8.3 & 29 & 24 & 21 & 11 \\\hline
GRB070208 & 4.4 & 5.5 & 16 & 29 & 26 & 15 \\\hline
GRB070219 & 14 & 21 & 59 & 16 & 14 & 8.0 \\\hline
GRB070223 & 13 & 16 & 47 & 17 & 15 & 8.9 \\\hline
GRB070318 & 4.9 & 6.1 & 21 & 28 & 25 & 13 \\\hline
GRB070330 & 3.7 & 5.5 & 16 & 32 & 26 & 15 \\\hline
GRB070412 & 5.9 & 8.8 & 25 & 25 & 21 & 12 \\\hline
GRB070420 & 25 & 22 & 74 & 12 & 13 & 7.1 \\\hline
GRB070427 & 4.4 & 5.5 & 16 & 29 & 26 & 15 \\\hline
GRB070506 & 15 & 22 & 54 & 16 & 13 & 8.4 \\\hline
GRB070508 & 6.1 & 9.1 & 26 & 25 & 20 & 12 \\\hline
GRB070509 & 7.9 & 12 & 34 & 22 & 18 & 11 \\\hline
GRB070529 & 9.0 & 11 & 32 & 20 & 18 & 11 \\\hline
GRB070611 & 3.5 & 4.4 & 15 & 33 & 29 & 16 \\\hline
GRB070621 & 4.6 & 4.7 & 16 & 29 & 28 & 15 \\\hline
GRB070714B & 46 & 69 & 160 & 9.0 & 7.4 & 4.8 \\\hline
GRB070721B & 9.6 & 14 & 41 & 20 & 16 & 9.6 \\\hline
GRB070805 & 8.0 & 14 & 34 & 22 & 16 & 10 \\\hline
%
  \end{tabular}
  \caption{
    Summary of fluence and distance constraints for waveforms a, b, and c.
    Distance limits are calculated assuming $E_\text{GW}=0.1 M_\odot c^2$.
  }
  \label{tab:ul}
\end{table*}

\section{Directional sensitivity cut}\label{dircut}
The pipeline used in this analysis works by comparing $\hat{Y}(t;f)$ (Eq.~\ref{eq:Y}) to $\sigma(t;f)$ (Eq.~\ref{eq:sigma}), which depends on the auto-power in the four neighboring segments on each side of $t$; (for additional details see~\cite{stamp}).
For a fixed direction on the sky $\hat\Omega$, the expectation value of $\text{SNR}(t;f)\equiv \hat{Y}(t;f)/\sigma(t;f)$ for one such pixel depends on the ``pair efficiency'' $\epsilon_{IJ}$ for each detector pair
\begin{equation}
  \langle \text{SNR}(t;f) \rangle \propto
  \frac{
    \epsilon_{12}(t;\hat\Omega)
  }{
    \left[\epsilon_{11}(t;\hat\Omega)\epsilon_{22}(t;\hat\Omega) \right]^{1/2}  } .
\end{equation}
Pair efficiency is defined in terms of the antenna response factors (see, e.g.,~\cite{stamp})
\begin{equation}
  \epsilon_{IJ}(t;\hat\Omega) \equiv \frac{1}{2} 
  \sum_A F^A_I(t,\hat\Omega) F^A_J(t,\hat\Omega) .
\end{equation}

For a small subset of directions $\hat\Omega$, the following condition is met
\begin{equation}
  \epsilon_{12} \ll 
  \left(\epsilon_{11}(t;\hat\Omega)\epsilon_{22}(t;\hat\Omega) \right)^{1/2} ,
\end{equation}
which means the GW signal produces a much stronger auto-power spectra compared to the cross-power spectra.
In the most extreme cases, the GW signal in the segments neighboring $t$ causes $\sigma(t;f)\gg \hat{Y}(t;f)$, which makes $\text{SNR}(t;f)\approx0$ even for loud signals.

To avoid searching in directions for which we are blind to GWs and can therefore not set limits, we employ a cut that ensures that GW signals can produce seed pixels with $\text{SNR}(t;f)\gtrsim1$:
\begin{equation}
  \epsilon_{12} \geq
  \frac{1}{4}
  \left(\epsilon_{11}(t;\hat\Omega)\epsilon_{22}(t;\hat\Omega) \right)^{1/2} .
\end{equation}
We find that this cut eliminates GRB triggers for which we cannot set effective limits while removing only one out of $51$ GRB triggers.

\end{appendix}

\bibliography{lgrb}

\end{document}